\documentclass[twocolumn, final]{svjour3}          
\usepackage{cite}
\usepackage{graphicx}
\usepackage[squaren, Gray, cdot]{SIunits}
\usepackage{amsmath}
\begin{document}

\title{Absolute airborne gravimetry with a cold atom sensor
}


\author{Yannick Bidel\and Nassim Zahzam \and Alexandre Bresson \and C\'edric Blanchard \and Malo Cadoret \and Arne V. Olesen \and Ren\'e Forsberg}  

\institute{Yannick Bidel  \and Nassim Zahzam \and Alexandre Bresson \and C\'edric Blanchard
\at DPHY,ONERA, Universit\'e Paris Saclay, F-91123 Palaiseau, France \\\email{yannick.bidel@onera.fr}
\and
Malo Cadoret
\at LCM-CNAM, 61 rue du Landy, 93210, La Plaine Saint-Denis, France
\and
Arne V. Olesen \and Ren\'e Forsberg
\at National Space Institute, DTU Space, Geodynamics Department, Kongens Lyngby, 2800, Denmark}




\maketitle

\begin{abstract}
Measuring gravity from an aircraft is essential in geodesy, geophysics and exploration. Today, only relative sensors are available for
airborne gravimetry. This is a major drawback because of the calibration and drift estimation procedures which lead to important operational constraints and measurement errors. Here, we report an absolute airborne gravimeter based on atom interferometry. This instrument has been first tested on a motion simulator leading to gravity measurements noise of 0.3 mGal for 75 s filtering time constant. Then, we realized an airborne campaign across Iceland in April 2017. From a repeated line and crossing points, we obtain gravity measurements with an estimated error between 1.7 and 3.9 mGal. The airborne measurements have also been compared to upward continued ground gravity data and show differences with a standard deviation ranging from 3.3 to 6.2 mGal and a mean value ranging from -0.7 mGal to -1.9 mGal.

\end{abstract}

\section{Introduction}
\label{intro}
Airborne gravimetry \cite{Forsberg2010} is a powerful tool for regional gravity mapping. It is relatively cheap, can cover large areas in a relatively short time and has good spatial resolution (around 5 km). Airborne gravimetry is especially interesting in the coastal areas where satellite altimetry does not work or over land areas which are difficult to access with terrestrial gravimetry (mountain areas, glaciers, deserts ...). 

Currently airborne gravity surveys are carried out with relative sensors \cite{Forsberg2015,Verdun2019,Jensen2019,Studinger2008} which can only measure the variation of gravity and which suffer from drift. For a gravity survey, one needs thus to go regularly to a reference point where the gravity is known or where a static absolute gravimeter is located. Additionally, the flight path design requires cross-over tracks, which are used in classical airborne gravimetry to determine drift parameters and signal validation. Therefore, the use of a relative gravimeter has important operational constraints which increase the time and the cost of gravimetry surveys. 

Two technologies exist for absolute gravimeter : optical and atomic. In optical gravimeters, the acceleration of a free falling corner cube is measured with optical interferometry \cite{Niebauer95}. These instruments are commercially available and can be operated only in static conditions. For dynamic operation, only one feasibility study done with a modified FGL gravimeter on an aircraft can be found in the literature \cite{Baumann12}. In an atom gravimeter, gravity is obtained from the acceleration measurement of a gas of cold atoms using matter wave interferometry \cite{Peters2001}. This latest technology has now reached or surpassed the performance of optical gravimeter \cite{Gillot2014, Freier2016,Hu2013} and start to be commercialized \cite{Menoret2018}. Moreover atom technology seems more adapted to dynamic environments because there is no mechanical moving parts and the repetition rate is higher. Recently, absolute ship borne gravimetry with sub-mGal precision has been reported using an atom gravimeter \cite{Bidel2018}.  The precision of the atom gravimeter called GIRAFE has been compared to a commercial spring gravimeter and showed better performances during the marine gravity campaign. 

Here, we report absolute airborne gravimetry with the GIRAFE atom gravimeter previously tested on a ship. In the first part, the atom gravimeter will be shortly described and the modifications compared to the previous marine test will be reported. In the second part, the airborne gravity campaign done in Iceland will be described. In the third part, the data processing to estimate gravity disturbance will be explained. Then, the results of the airborne campaign will be shown. Finally, in the last part, the airborne measurements will be compared with ground data.

\section{Cold atom gravimeter}

\subsection{Apparatus description}
The description of the gravimeter can be found in the reference \cite{Bidel2018} and we provide here only a short description. The gravimeter is composed of an atom sensor which provides an absolute measurement of the acceleration, a gyro-stabilized platform which maintains the accelerometer aligned with the local gravity acceleration despite angular movements of the carrier and systems which provide the lasers and
microwaves needed to the atom sensor and perform data acquisition and processing.

\begin{figure}[ht!]
	\begin{center}
	\centerline{\scalebox{0.42}{\includegraphics{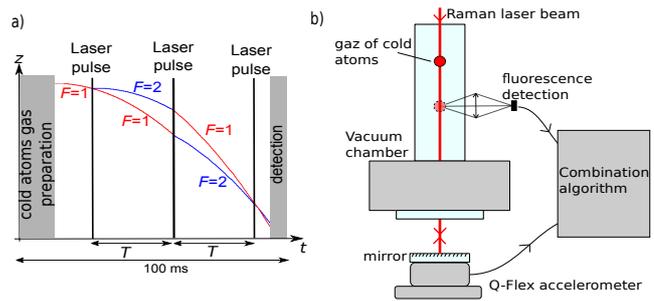}}}
	\end{center}
\caption{Principle of the atom accelerometer. a) Temporal sequence. b) Set-up of the atom accelerometer.}
\label{atom}
\end{figure}

The principle of the atom accelerometer is based on the acceleration measurement of a free falling test mass. The test mass is a gas of cold Rubidium 87 atoms produced by laser cooling and trapping method. The trapped gas contains typically $10^6$ atoms, has a size of  1 mm and a temperature of 1 \micro K. After release from the trap, atoms are let in free fall and their accelerations are measured by an atom interferometry. For that, the atoms are submitted to three laser pulses separated by a duration T. The laser pulses drive two photon Raman transitions between the two hyperfine ground states of the atoms and give a momentum to the atoms when they undergo the transition. The first pulse acts as a matter wave beam splitter, the second one acts as a mirror and the last one recombines the matter waves (see Fig. \ref{atom}). The signal of the atom interferometer is then obtained by measuring the proportion of atoms in the two hyperfine states by laser induce fluorescence method. The output P of the atom sensor is proportional to the cosine of the acceleration with a period equal to $\lambda/2T^2$ where $\lambda=780$ nm is the laser wavelength. In our sensor the pulse separation T can be changed. Our 14 mm falling distance allows us to change T from 0 to 20 ms. For T = 20 ms, the period is equal to \unit{10^{-3}}{\meter.\rpsquare \second} and is small compared to typical variations of acceleration in a moving vehicle. There is, therefore, an ambiguity to determine the acceleration from the measurement of the atom sensor. Many values of acceleration are possible for a given value of the output of the atom sensor. To overcome this limitation, we combine the atom sensor with a force balanced accelerometer (Qflex from Honeywell). The classical accelerometer is used to give a first rough estimation of the acceleration in order to determine which value of acceleration corresponds to the signal of the atom sensor. The classical accelerometer is also used to measure the acceleration during the measurement dead times of the atom sensor which occur during the cold atoms preparation and during the detection. On the other hand, the atom accelerometer allows to estimate the bias of the classical accelerometer and thus improving its precision. 

This hybridization is working if the difference of acceleration given by the two sensors is much smaller than the atom accelerometer signal period ($\lambda/2T^2$). Different limitations can induce differences of acceleration and specially in hard dynamical environments (transfer function uncertainties, alignment defaults, measurement points non co-located). In order to be always operational, the gravimeter algorithm is changing automatically the atom interrogation time T (T = 2.5, 5, 10 or 20 ms) by comparing the rms on the difference of acceleration given by the two sensors and the atom accelerometer period. If the rms difference is small, the algorithm will increased the interrogation time and the gravimeter will thus access to better precision  due to the scale factor increase. If the rms difference is too big, the algorithm will decrease the interrogation time T which will allow the gravimeter to keep working but this will also decrease the precision measurement. During the different tests describe in this article, the interrogation time will stay at T=20 ms excepted during turbulent parts of flight where the interrogation time switches to T = 10 ms.

This atom accelerometer has been implemented in a compact housing consisting of a cylinder of
22 cm diameter and 52 cm height. It is composed of a vacuum chamber made of glass in which the atoms are
produced and interrogated, magnetic coils, optics for shaping all the laser beams and collecting the
fluorescence of the atoms, two layers of mu-metal for shielding the external magnetic field and classical
accelerometers. This sensor is integrated in a two axes stabilized gimbaled platform made by IMAR. The
platform is stabilized using an integrated inertial measurement system and maintains the sensor head aligned
with the gravity acceleration with a precision of 0.1 mrad. The platform is mounted on passive vibration isolators which have a resonant frequency of 12 Hz.

In static condition, the sensitivity of the gravimeter is equal to 0.8 mGal$\cdot$Hz$^{-1/2}$ and the accuracy is estimated at 0.17 mGal for T=20ms \cite{Bidel2018}. 

\begin{figure*}[ht!]
	\begin{center}
	\centerline{\scalebox{0.4}{\includegraphics{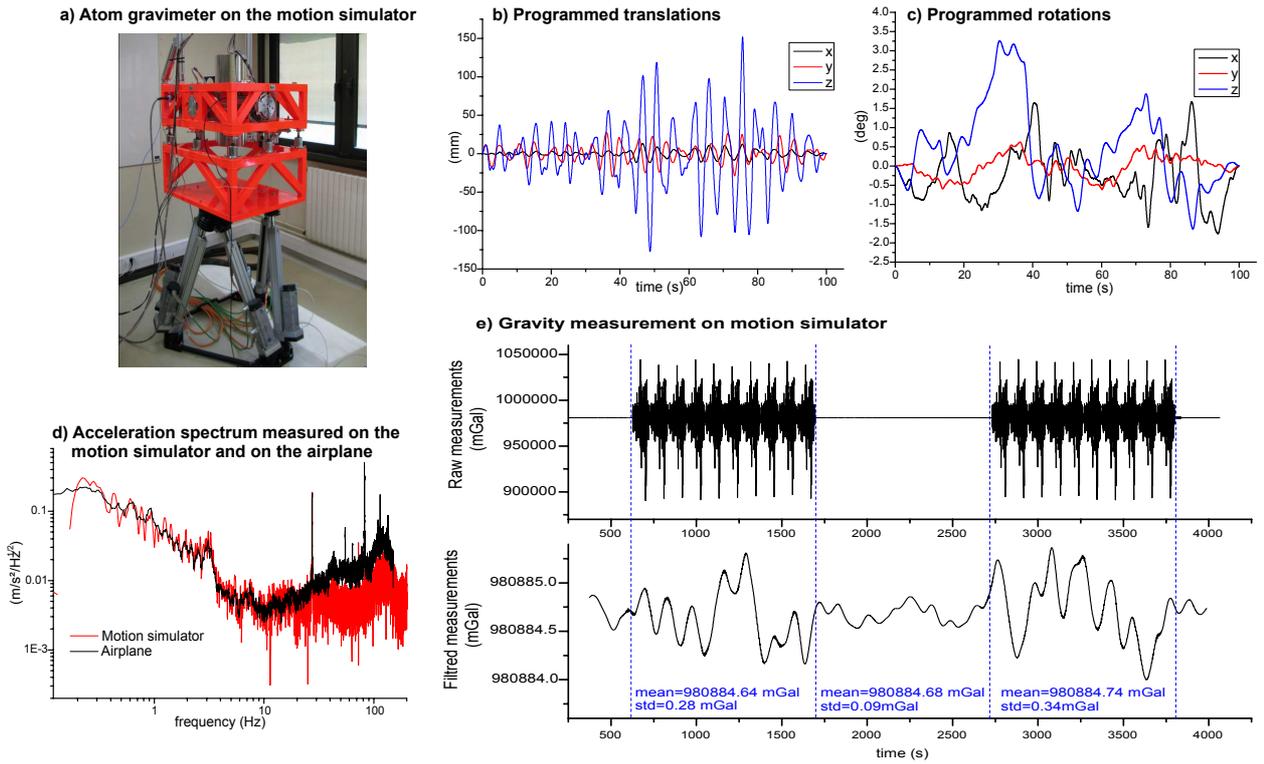}}}
	\end{center}
\caption{Test on a motion simulator. a) Picture of the atom gravimeter on the motion simulator.  b) Programmed translation on the motion simulator along the three axes. c) Programmed rotation on the motion simulator along the three axes. d) Vertical acceleration spectrum measured on the motion simulator (red) and on the real flight (black). e) Gravity measurement on the motion simulator (top : raw data, bottom, filtered data with 4th order Bessel filter of time constant 75 s)}
\label{MS}
\end{figure*}

\subsection{Improvement of the force balanced accelerometer model for high frequency vibrations}

In airborne environment, the gravimeter is subjected to strong vibrations. In this case, if we do not take into account the exact transfer function of the force balanced accelerometer, the acceleration given by the atom and the force balanced accelerometer could be different and not negligible compared to the period of the atom accelerometer signal (\unit{10^{-3}}{\meter.\rpsquare \second} for $T=$20 ms). In this situation, the hybridization method will not work properly and will lead to decrease of performance of the gravimeter. The transfer function of the force balanced accelerometer has thus to be known precisely and compensated in order to optimize the precision of our instrument. 

The transfer function of our force balanced accelerometer (Qflex) has been estimated empirically by minimising the difference between the  acceleration from the force balanced accelerometer and the atom accelerometer in presence of high frequency vibrations. For that, we model the transfer function of the force balanced accelerometer by a first order damped harmonic oscillator:
\begin{equation}
h_{\text{FB}}(s)=\frac{\omega_0^2}{s^2+\Gamma s+\omega_0^2}\;\;\text{;}\;\;s=j\omega
\end{equation}
We obtained for the parameters of the transfer function $\omega_0 = 1.57 \cdot 10^{3}\; \text{s}^{-1}$ and $\Gamma=2.42\cdot 10^3\; \text{s}^{-1}$.

\subsection{Test on a motion simulator}

The atom gravimeter has been tested on a motion simulator reproducing as well as possible the motion of an aircraft (see Fig. \ref{MS} a).
For that, we took 100 s of IMU data coming from a DTU flight campaign in Antarctica with a Twin-Otter (non-turbulent part). Then we programmed the motion simulator to reproduce the three translations and three rotations measured by the IMU. The translations were high pass filtered at a frequency of 0.2 Hz for having translation in the range of the motion simulator ($\pm$ 0.18 m).

To check the fidelity of the simulation, we measured the vertical acceleration on the base plate of the
gravimeter and we compared it with the acceleration coming from the IMU of the plane. We notice that the motion simulator reproduced well the acceleration spectrum between 0.2 Hz and 20 Hz (see Fig. \ref{MS} d). 

The gravimeter was subjected to a simulated airborne environment during two periods of 1000 s with a
break of 1000 s between them (see Fig. \ref{MS} e). The gravimeter measurement were low pass filtered by a 4th order Bessel filter of 75 s time constant (see \ref{Lowpass}). We notice that the mean value of measured gravity has not significantly
changed during the period of motion simulation. The rms noise on the filtered gravity measurement is equal
to 0.3 mGal during motion and 0.1 mGal during static period.

\section{Airborne gravity campaign in Iceland}

\begin{figure}[ht!]
	\begin{center}
	\centerline{\scalebox{0.32}{\includegraphics{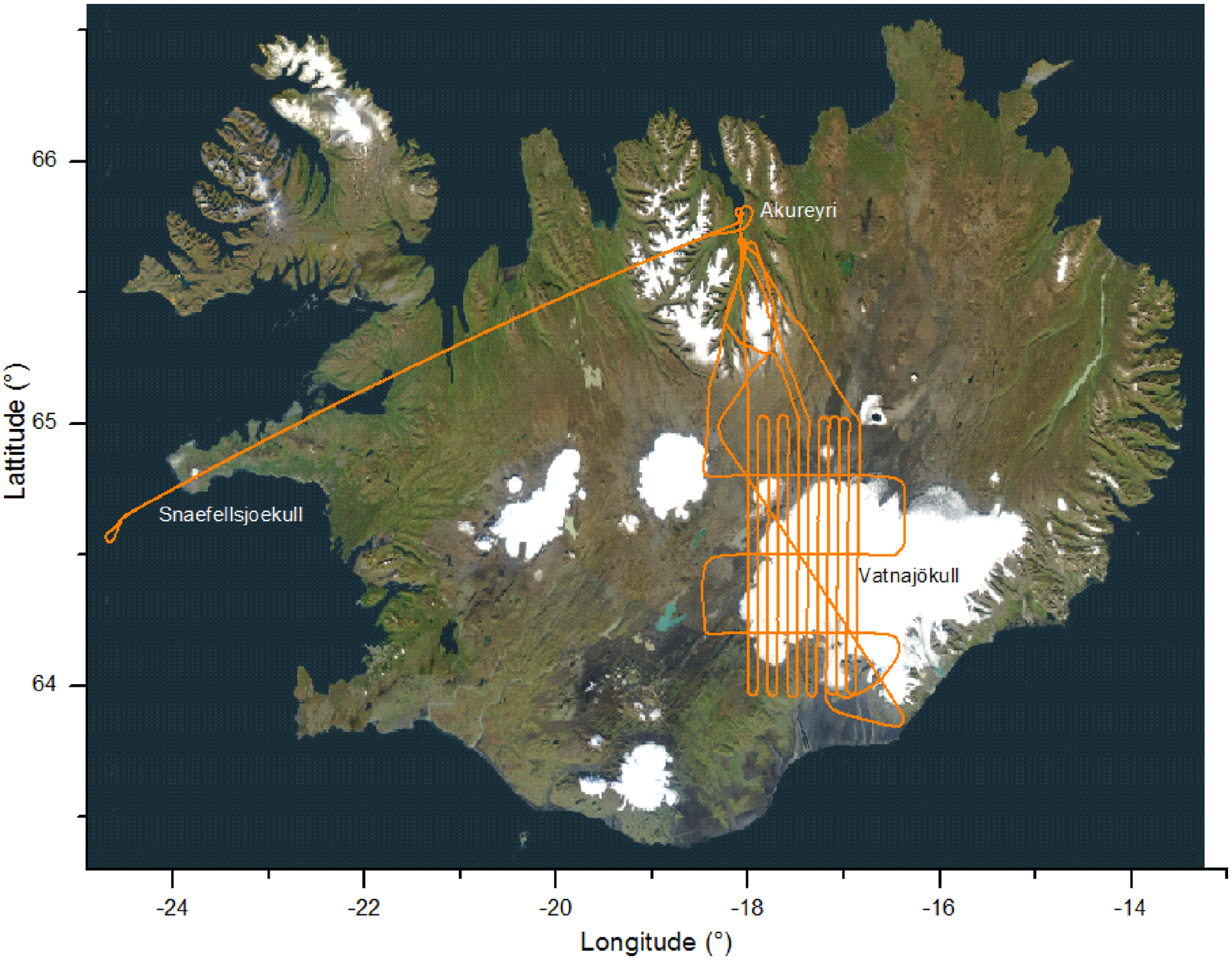}}}
	\centerline{\scalebox{0.29}{\includegraphics{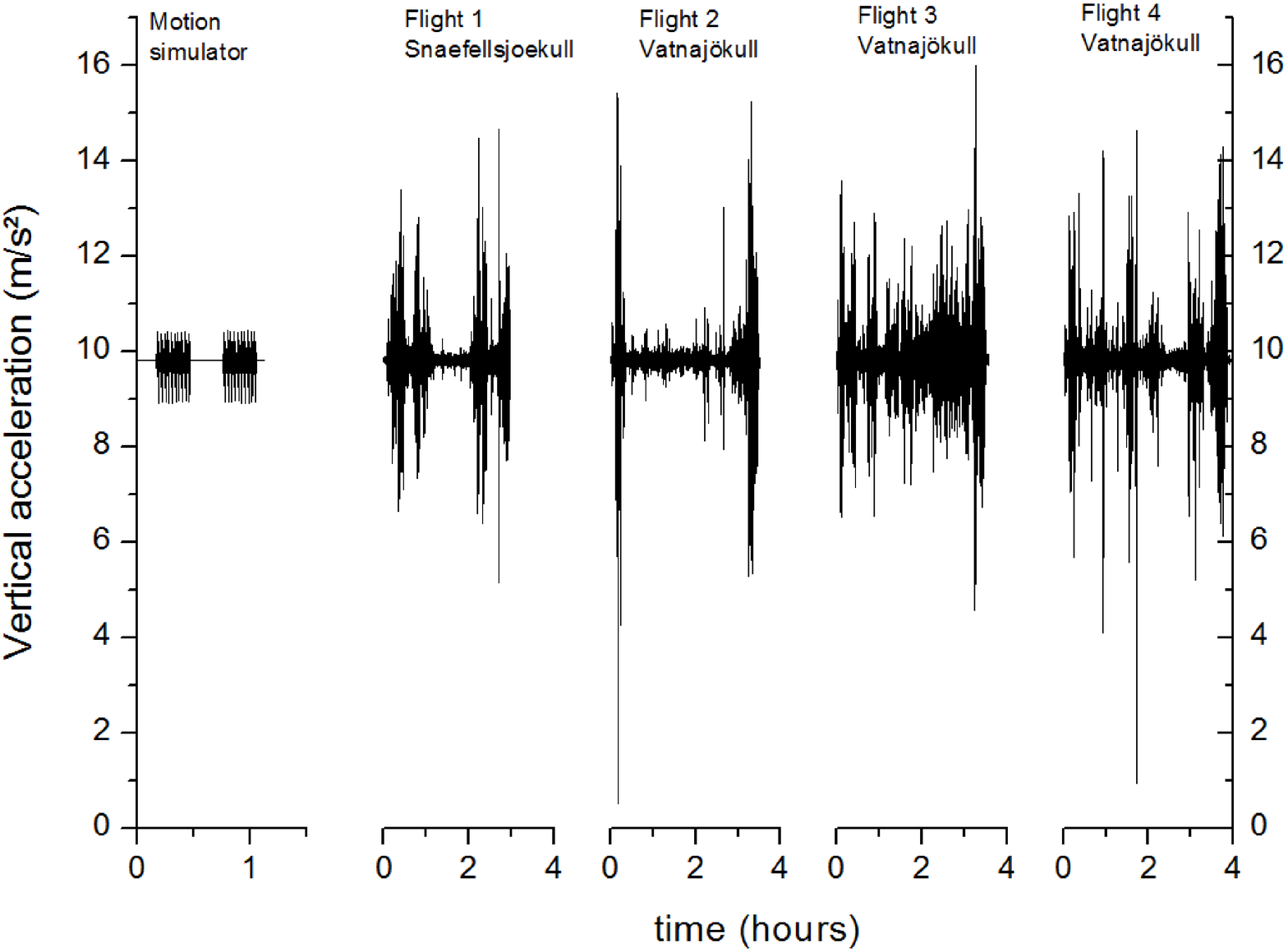}}}
	\end{center}
\caption{Top: Flight plan of Iceland gravity campaign. Bottom: Raw vertical acceleration undergone by the atom gravimeter during the motion simulator test and during flights in Iceland. The acceleration has been measured in the sensor head at a rate of 10  Hz.}
\label{flights}
\end{figure}

The campaign took place across Iceland, using a Twin Otter DHC-6 from Norlandair (Akureyri) and consisted of repeat flights in northern Iceland and a small demonstration survey pattern over the Vatnaj\"okull (see Fig. \ref{flights}).

Before airborne tests, we performed static measurement in the plane hangar. We obtained a gravity measurement of g = 982 337.37 $\pm$ 0.17 mGal at 99 cm above the ground which agrees with a previous measurement made with a A10 absolute gravimeter to within 0.1 mGal. 

The atom gravimeter was tested during four flights: the first one was a straight line back and forth between Akureyri and Sn\ae fellsj\"okull. The goal of this flight is to evaluate the reproducibility of the gravity measurement. The last three measurement flights were above Vatnaj\"okull. The goal was here to make a gravity model of the area. The duration of each flight was 3 - 4 hours.
The vertical acceleration measured during the flights is given on Figure \ref{flights}. We notice that the acceleration level during the flights is not homogeneous. During turbulent part, one can have acceleration variations up to \unit{10}{\meter.\rpsquare \second} and during quiet part  below \unit{0.3}{\meter.\rpsquare \second}. We notice also that most of the time the level of acceleration is larger than the one we simulated on the motion simulator.

\section{Data processing and gravity estimation}

\subsection{Kinematic acceleration and E\"otv\"os effect}

The gravimeter is not only measuring the gravity acceleration but also the kinematic acceleration of
the plane and the acceleration due to the coupling to Earth rotation (E\"otv\"os effect). The acceleration
measured by the gravimeter is equal to :
\begin{equation}
a_{\text{meas}}=g+\ddot{h}+a_{\text{E\"ot}}
\end{equation}
where g is the gravity acceleration, $\ddot{h}$ is the time second derivative of the altitude and represents the vertical
kinematic acceleration of the plane, $a_{\text{E\"ot}}$ is the E\"otv\"os acceleration which is equal to :
\begin{equation}
a_{\text{E\"ot}}=-2\,\omega_E\cdot \cos(\varphi)\cdot v_{E}-\frac{v_{E}^2}{N(\varphi)+h}-\frac{v_{N}^2}{M(\varphi)+h}
\end{equation}
with:
\small
\begin{equation}\nonumber
\begin{array} {lll}
\omega_E=7.292115\cdot 10^{-5}\,\text{s}^{-1}&:&\textnormal{\small Earth's rotation rate  }\\
&&\textnormal{\small (inertial frame) }\\
\\
\varphi &:&\textnormal{\small Latitude}\\
\\
v_{E}&:&\textnormal{\small East velocity}\\
\\
v_{N}&:&\textnormal{North velocity}\\
\\
h&:&\textnormal{\small Altitude}\\
\\
M(\varphi)=&:& \textnormal{\small Earth's radius of curvature}\\
\frac{a^2\cdot b^2}{\left(a^2\cos(\varphi)^2+b^2\sin(\varphi)^2\right)^{3/2}}&&\textnormal{in the (north-south) meridian}\\
\\
N(\varphi)=&:& \textnormal{\small Earth's radius of curvature}\\
\frac{a^2}{\left(a^2\cos(\varphi)^2+b^2\sin(\varphi)^2\right)^{1/2}}&&\textnormal{in the prime vertical}\\
\\
a=6378137.0\, \meter&:&\textnormal{\small Earth's equatorial radius}\\
&&\textnormal{\small (WGS84)}\\
\\
b=6356752.3\,\meter&:&\textnormal{\small Earth's polar radius}\\
&&\textnormal{\small (WGS84)}\\
\end{array}
\end{equation}
\normalsize

The vertical kinematic acceleration and E\"otv\"os effect are calculated with GNSS data ($\varphi$: latitude, $\lambda$:
longitude, $h$: altitude) at 10 Hz ($dt=$0.1 s) based on differential and post-treated DGPS data. The level arm between the GNSS antenna and the gravimeter has been taken into account. The vertical kinematic acceleration, the east velocity and the north velocity have been calculated using the
following equations:
\begin{equation}
\begin{array} {lll}
\ddot{h}(t)&=&\frac{-2\,h(t)+h(t+dt)+h(t-dt)}{dt^2}\\
\\
v_{E}(t)&=&\frac{\lambda(t+dt)-\lambda(t-dt)}{2\,dt}\cdot\left(N(\varphi)+h\right)\cdot \cos(\varphi)\\
\\
v_{N}(t)&=&\frac{\varphi(t+dt)-\varphi(t-dt)}{2\,dt}\cdot\left(M(\varphi)+h\right)
\end{array}
\end{equation}

\subsection{Missing data points and interpolation}

The gravimeter provides acceleration measurements at a rate of 10 Hz. The precise timing of the
 measurements compared to the GNSS is crucial in order to correct precisely from the effect of
kinematic acceleration and E\"otv\"os effect which can be up to $10^6$ times bigger than the gravity disturbance
signal. However, the timing of the gravimeter measurements is not precise and has the following default:\\
- the clock of the computer which controls the gravimeter is not precise (relative drift of ~ $3\cdot 10^{-5}$) and
has an unknown delay compared to the GNSS time base;\\
- the recording time has jitters compared to the real measurement time of the gravimeter;\\
- there are missing data points (typically 1 per hour);\\
- there is a 20 ms offset of the effective measurement time compared to the recording measurement
time when the interrogation time $T$ of the gravimeter is changing between 10 ms and 20 ms.\\

We try to correct these limitations by using the following procedure. First, the missing data points are filled by inserting extrapolated measurements. Second, we assume that the measurement times of the gravimeter are given by : $t_i = i. dt +T+
t_0$ where $dt\sim 0.1$ s is the time interval between measurements and $T$ is the interrogation time used by the
gravimeter. Then, we adjust the parameter $dt$ and $t_0$ in order that the acceleration given by the GNSS and the
gravimeter match at the beginning and at the end of the acquisition period.

\subsection{Lowpass filtering}\label{Lowpass}
The gravimeter measurement, the kinematic acceleration and the E\"otv\"os effect are filtered with a
4th order Bessel low pass filter of time constant $\tau$ = 130 s :
\begin{equation}
h(s)=\frac{105}{s^4+10s^3+45s^2+105s+105}\;\;\text{;}\;\;s=j\omega\tau
\end{equation}
For a plane of velocity v, this gives a spatial resolution equal to $\approx 1.035\cdot v \cdot \tau$. The spatial resolution is here defined as the FWHM of the signal obtained with a Dirac input signal. For the filter to work properly, we linearly extrapolate the gravity measurements points and the GNSS data on a regular time base at 10 Hz.

\subsection{Gravity disturbance calculation}

The gravity disturbance is obtained by subtracting the gravity measurements by the WGS84 normal gravity model
taking into account altitude and latitude effects \cite{Torge89}:
 \begin{align}  
g_0=\frac{a\cdot g_E\cdot \cos(\varphi)^2+b\cdot g_P\cdot \sin(\varphi)^2}{\sqrt{a^2\cdot \cos(\varphi)^2+b^2\cdot \sin(\varphi)^2}}
\cdot (1+\gamma_1\cdot h\nonumber \\
+\gamma_2 \cdot h^2 ) 
\end{align}
with :
 \small
\begin{equation}
\begin{array}{l}
g_E=9.7803253359\,\text{m}\cdot\text{s}^{-2}\;\;\text{(WGS84)} \\
\\
g_P=9.8321849378\,\text{m}\cdot\text{s}^{-2}\;\;\text{(WGS84)}\\
\\
\gamma_1=-\frac{2}{a}\left(1+f+\frac{a^2\cdot b\cdot \omega_E^2}{G.M}-2\cdot f \cdot \sin(\varphi)^2\right)\\
\\
\gamma_2=\frac{3}{a^2}\\
\\
f=\frac{a-b}{a} \\
\\
G.M = 3.986004418\cdot 10^{14}\, \text{m}^3\cdot \text{s}^{-3}\;\;\text{(WGS84)}
\end{array}
\end{equation}
\normalsize

\subsection{Correction of the alignment errors of the platform}

Alignment errors of the platform make the gravimeter less sensitive to vertical gravity acceleration and make it
sensitive to horizontal accelerations. To evaluate this error, we follow the modelling approach described in the
thesis of A.V. Olesen \cite{Olesen02}.
The error on gravity measurements caused by a platform misalignment is given by:
\begin{equation}
\delta g_{\text{tilt}}=\frac{\phi_x^2+\phi_y^2}{2}\cdot g +\phi_x\cdot a_x+\phi_y\cdot a_y
\end{equation}
where $\phi_x$  and $\phi_y$  are the misalignment angle compared to the direction of the gravity acceleration and $a_x$ and
$a_y$ are the horizontal accelerations. In this expression, we assume that the misalignment angles are small ($\phi_x$, $\phi_ y << 1$). The misalignment angles are estimated by comparing the accelerations measured by horizontal force balance accelerometers located in the sensor head and the kinematic acceleration deduced from GNSS data:
\begin{equation}
\phi_{x(y)}=\frac{a_{x(y)}-a_{x(y)\text{\tiny GNSS}}}{g}
\end{equation}
The parameter $a_x$, $a_y$, $a_{x\text{GNSS}}$ and $a_{y\text{\tiny GNSS}}$ have been pre-filtered by a 4th order Bessel filter of time
constant 40 s. The correction tilt $\delta g_{\text{tilt}}$ obtained has been filtered with the same filter than the
gravimeter measurement i.e. a 4th order Bessel filter with a time constant of 130 s.
We obtained alignment errors up to 20 mGal in period of gravity measurements i.e. constant yaw.
This error is very different from flight to flight (see Table \ref{errplat}).
\begin{table}
\caption{Error from platform misalignment}
\label{errplat}       
\begin{tabular}{lll}
\hline\noalign{\smallskip}
 & $\delta g_{\text{tilt}}$ max\\
\noalign{\smallskip}\hline\noalign{\smallskip}
Flight 1 : Akureyri-Snaefellsjokull& 1 mGal \\
Flight 2 : Vatnajokull & 20 mGal \\
Flight 3 : Vatnajokull & 4 mGal\\
Flight 4 : Vatnajokull & 5 mGal\\
\noalign{\smallskip}\hline
\end{tabular}
\end{table}

\section{Airborne test results}

\subsection{Akureyri-Sn\ae fellsj\"okull}

\begin{figure}[ht]
	\begin{center}
	\centerline{\scalebox{0.295}{\includegraphics{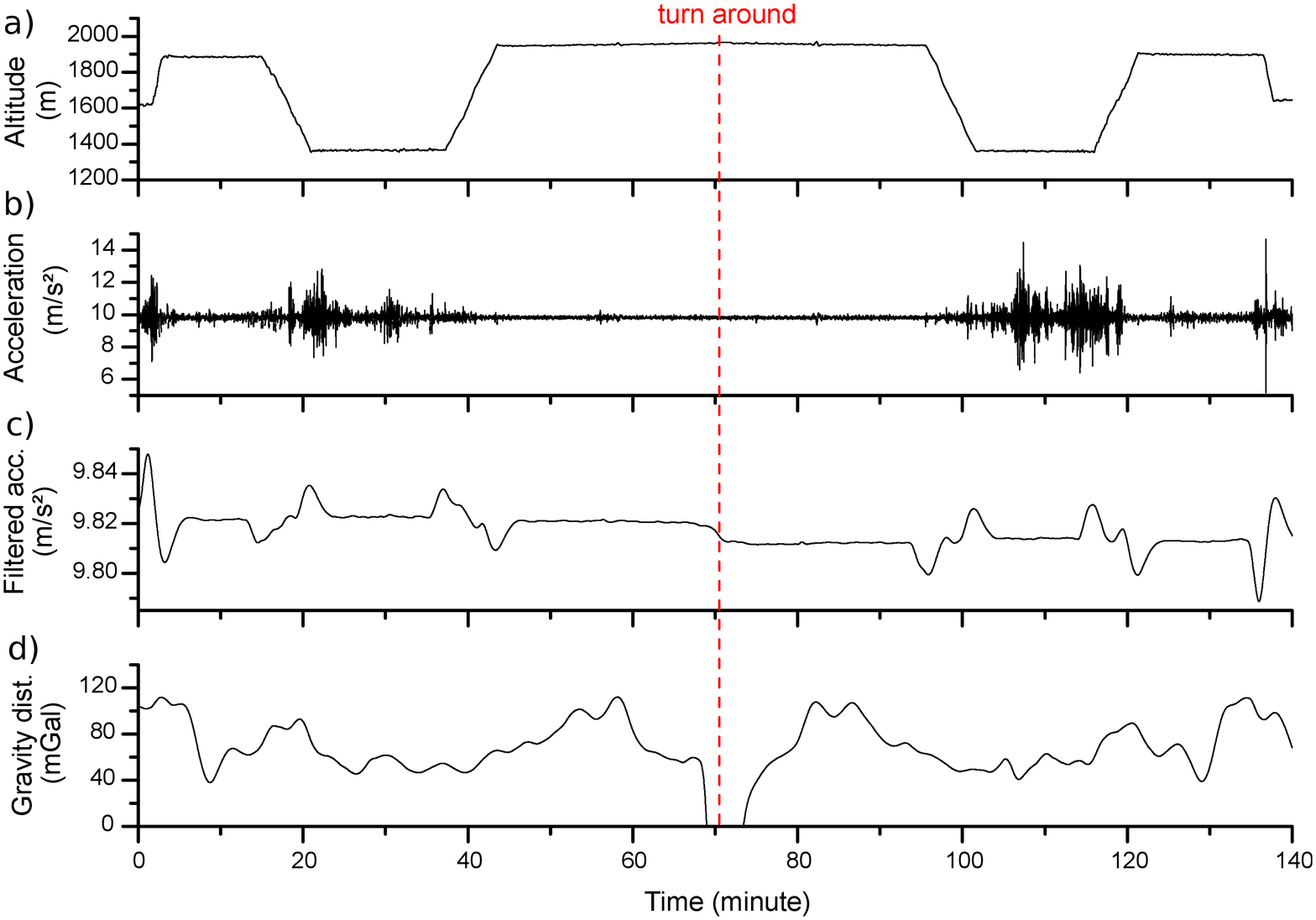}}}
	\end{center}
\caption{Gravity measurements on the Akureyri Sn\ae fellsj\"okull line. a): Altitude of the plane. b) Raw acceleration measured by
the gravimeter. c) Filtered acceleration measured by the gravimeter (4th order low pass Bessel filter of time constant 130 s). d)
Estimated gravity disturbance with the 130 s low pass filter.}
\label{akur}
\end{figure}

\begin{figure}[ht]
	\begin{center}
	\centerline{\scalebox{0.28}{\includegraphics{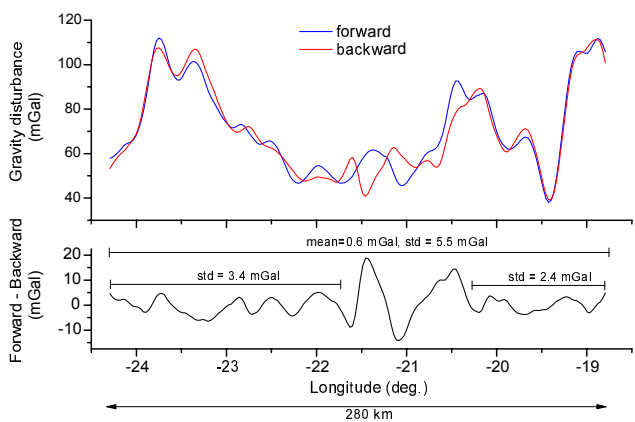}}}
	\end{center}
\caption{Comparison of the gravity measurement along the line Akureyri- Sn\ae fellsj\"okull for the forward and backward flight.}
\label{AkSn_comp}
\end{figure}

The airborne measurements obtained on the line Akureyri - Sn\ae fellsj\"okull flown back and forth are given on
Fig. \ref{akur}. The plane was flying at two elevations (1900 m and 1400 m) in order to be as close as possible to
the ground and thus to the gravity sources. The 1900 m altitude corresponds to mountain area and the
1400 m elevation corresponds to plain area. The velocity of the plane was 76 m/s. With the 4th order Bessel
filter of time constant 130 s, one obtains a spatial resolution of 10.5 km (FHWM). On the filtered acceleration graph,
one can see clearly the E\"otv\"os effect when the plane turned around. Indeed, at this point the velocity changes
of sign and the E\"otv\"os acceleration also. On can also see clearly the effect of the vertical acceleration of the plane
at the moment where the plane was changing of elevation.
In order to estimate the repeatability of the measurements, we compared the gravity measured
forward and backward (see Fig. \ref{AkSn_comp}). The difference between forward and backward has a mean of
0.6 mGal and a standard deviation of 5.5 mGal. One notices that the big difference in the centre corresponds
to some missing measurement points on the gravimeter measurements. If one restricts to the area where there
is no missing points, one obtains a standard deviation of 3.4 mGal close to Sn\ae fellsj\"okull and 2.4 mGal close
to Akureyri. Assuming uncorrelated errors between forward and backward measurements, the measurement
error is given by the standard deviation of the difference divided by $\sqrt{2}$ . One obtains thus an estimated error
ranging from 1.7 mGal to 3.9 mGal depending on the area considered.

\subsection{Vatnaj\"okull}

\begin{figure}[ht]
	\begin{center}
	\centerline{\scalebox{0.067}{\includegraphics{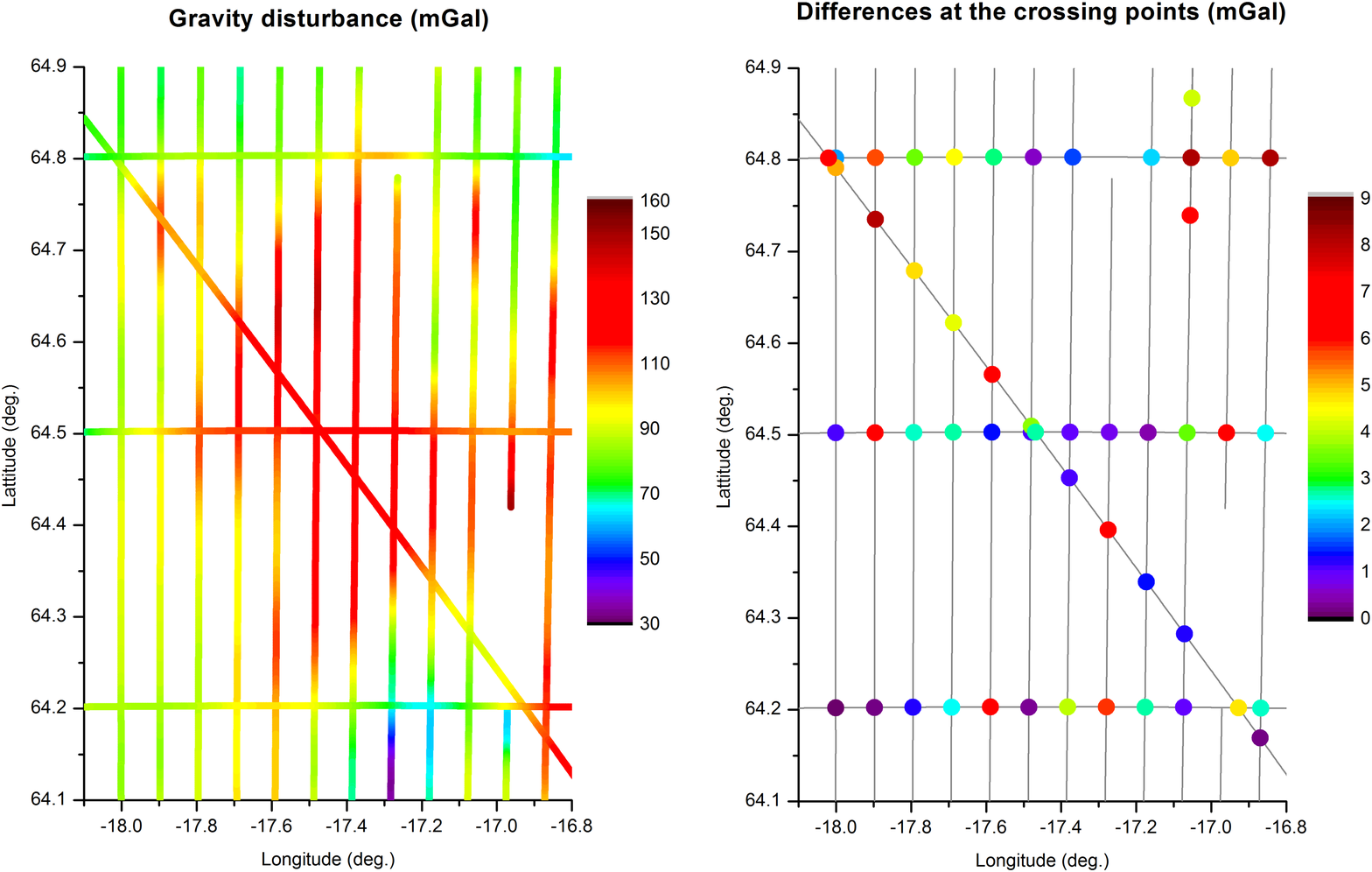}}}
	\end{center}
\caption{Vatnaj\"okull gravity measurements. Left: gravity disturbance. Right: Crossing points differences}
\label{Vat}
\end{figure}

During three flights, we measured gravity above the area of Vatnaj\"okull ice cap along 16 lines. The altitude of the plane was 2600 m and its velocity 76 m/s. We use the same filter than before leading to a spatial resolution of 10.5 km. The gravity disturbance measurements obtained are reported on Fig. \ref{Vat}. One notices two measurements area missing which correspond to moments where the gravimeter was not operational due to laser misalignment problems. The difference at the crossing points are ranging from 0 to 8 mGal with a rms value of 3.9 mGal. Assuming no correlation, one can estimate a measurement error of 2.8 mGal (rms value divided by  $\sqrt{2}$).

\section{Comparison with ground data}

\begin{figure}[ht]
	\begin{center}
	\centerline{\scalebox{0.62}{\includegraphics{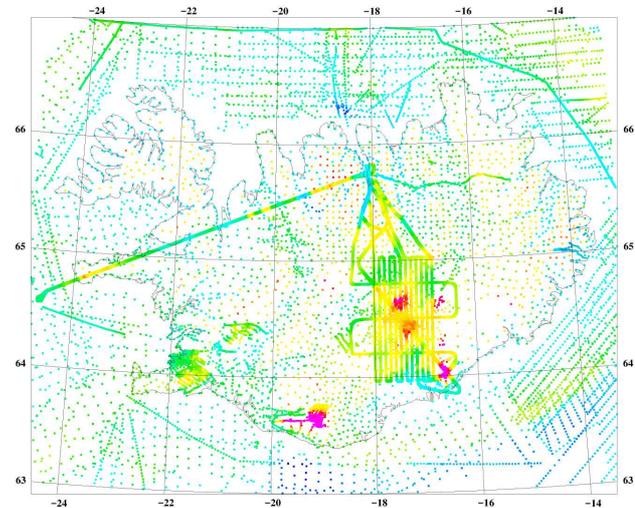}}}
	\end{center}
\caption{Iceland gravity coverage (ground measurements), overlaid with the cold atom gravimetry results. The positive free-air anomalies shown are predominantly due to volcanoes under the ice caps, and topographic highs.}
\label{ground}
\end{figure}

The Iceland region has a relatively dense ground gravity coverage, as shown in Fig. \ref{ground}. The use of upward continued surface gravimetry represents an independent validation opportunity for the cold atoms gravimetry results. The Iceland gravity data were surveyed primarily in the 1980’s, and provided by Landm\ae lingar Islands (Iceland Geodetic Survey).

The upward continuation estimation of the free-air anomalies at altitude were done using the GRAVSOFT suite of programs \cite{Tscherning1992}, using standard remove-restore techniques of physical geodesy (use of EGM2008 as reference field, integration of terrain effects by prism integration,  and upward continuation to the flight altitude by Fast Fourier transform methods \cite{Schwarz1990}). A digital terrain model at 200 m resolution was used and combined with a ice cap thickness model of the 3 main ice caps in Iceland (including Vatnaj\"okull), derived from radar echo soundingand also provided by Landm\ae linger Islands, as part of cooperation on geoid determination.

\begin{figure}[ht]
	\begin{center}
	\centerline{\scalebox{0.29}{\includegraphics{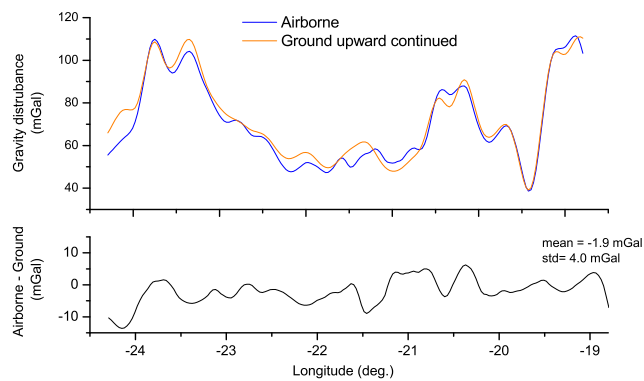}}}
	\end{center}
\caption{Comparison between airborne measurements (average of forward and backward)and ground measurements upward continued along the line Akureyri-Sn\ae fellsj\"okull}
\label{CompGroundLine}
\end{figure}

\begin{figure}[ht]
	\begin{center}
	\centerline{\scalebox{0.44}{\includegraphics{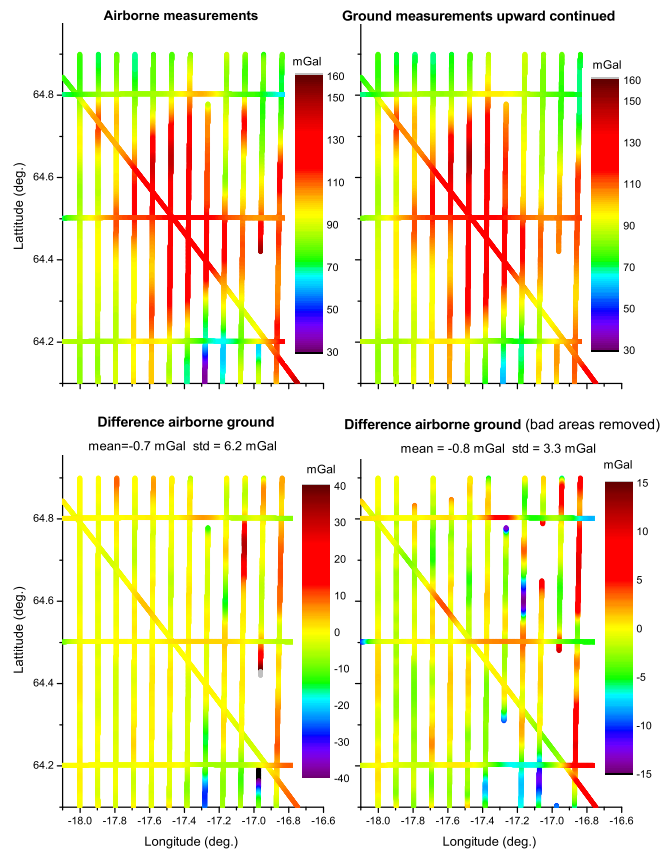}}}
	\end{center}
\caption{Comparison between airborne measurements and ground measurements upward continued over  Vatnaj\"okull.}
\label{CompGroundVat}
\end{figure}

The predicted versus the observed cold atom gravimetry results are shown in Fig. \ref{CompGroundLine} and Fig. \ref{CompGroundVat}, with the predicted data at altitude filtered with a similar 4th order Bessel filter with time constant 130 s, to match the airborne data filter. One notices that similar gravity signals are obtained with the two models confirming the relevance of the cold atom gravimeter measurements. For the line Akureyri-Sn\ae fellsj\"okull, we obtained a standard deviation on the difference equal to 4.0 mGal and a mean difference of -1.9 mGal; it should be noted that some part of this line was over fjords with no surface gravity, and the upward continued gravity data may therefore be biased.
For Vatnaj\"okull flights, we obtained a standard deviation on the difference equal to 6.2 mGal and a mean difference of -0.7 mGal. We noticed that in some areas (see Fig. \ref{CompGroundVat}), the difference between airborne and ground is large. This areas correspond to the beginning of a track (after a plane turn), to a period around laser misalignment problem and to a severe turbulence period ($\varphi$ = 64.7$^{\circ}$, $\lambda$ = -17.1$^{\circ}$). If we removed this areas, the standard deviation becomes two times smaller (3.3 mGal) and the mean difference is approximatelly the same (-0.8 mGal).

An issue for the comparison of surface and airborne data is also the possible geodynamic gravity changes between the surface and airborne gravity epochs, since several volcanic eruptions have taken plane, especially the Bardabunga eruption of 2014, which had major dyke intrusion activity in the northwestern region of the Vatnaj\"okull ice cap.

\section{Conclusion}

In conclusion, we demonstrated for the first time airborne gravity measurements and survey with an atom interferometry sensor. The main advantage of this technology is that it provides absolute measurements (no drift and no calibration needed). The precision of the gravity measurements have been estimated thanks to comparison on a forward and backward line and to differences at crossing points. Measurement errors ranging from 1.7 to 3.9 mGal have been obtained. The airborne gravity measurements have been also compared to upward continued ground truth. The standard deviation on the difference is ranging from 3.3 to 6.2 mGal and the mean value on the difference is ranging from -0.7 to -1.9 mGal.

This is a promising result for a sensor which was designed for marine application. The precisions obtained here could be improved by optimizing the instrument on the followings points :\\
- Improving the measurement timing of the atom gravimeter : measurements points on a regular time basis (GNSS dating).\\
- Suppressing the missing measurements points.\\
- Optimizing the gyro-stabilized platform.\\
- Optimizing the hybridization algorithm between the force balanced and the atom accelerometer for airborne environment.\\
With these improvements which are not inherent to atom interferometry technology, atom gravimeter should reach the state of the art with sub mGal precision on airborne survey with still absolute measurements.

Finally, these results show the maturity of cold atom technology for onboard application and support the development of atom interferometry sensor for measuring the Earth gravity field from space \cite{Carraz14,Abry2019}.

\begin{acknowledgements}
The development of the atom gravimeter was funded by the French Defense Agency (DGA). The Iceland GIRAFE cold atom airborne campaign was carried out with support from ESA (Cryovex), ONERA and DTU Space. Nordlandair, Iceland, provided excellent support for the challenging installation of the complex system in the Twin-Otter TF-POF. We thank Landm\ae lingar Islands for providing the permission at short notice for the test campaign in Iceland.

\end{acknowledgements}

\bibliographystyle{ieeetr}

\bibliography{biblio}   

%
%


\end{document}